\documentclass[aps,prl,twocolumn,amssymb,showpacs]{revtex4}
\begin{document}
\title{Is there a ``native'' band gap in ion conducting glasses?}
\author{Jeppe C. Dyre}
\affiliation{Department of Mathematics and Physics (IMFUFA),
Roskilde University, Postbox 260, DK-4000 Roskilde, Denmark}
\date{\today}

\begin{abstract}
It is suggested that the spectrum of ion site energies in glasses exhibits a
band gap, establishing an analogy between ion conducting glasses and intrinsic 
semiconductors. This implies that ion conduction in glasses takes place via 
vacancies and interstitial ions (as in crystals).
\end{abstract}
\maketitle

Ion conduction in glasses has been studied for many years but there is still no 
universally accepted theory 
\cite{ste57,owe63,hav65,isa69,moy81,ang90,fun90,bun98}. Some theories assume a 
more or less collective conduction mechanism, others assume that conduction 
proceeds via defects like vacancies or interstitials. We shall argue that most likely the 
latter is always the case. The approach taken below is to consider the very basic 
questions which may be asked.

The obvious first question relates to ion statics: 
\begin{itemize} 
\item {\it What is the nature of the states of the ionic system?}
\end{itemize}
It is quite clear that any glass has a discrete number of possible ion sites and that 
each ion site has room for just one ion (ions have a substantial volume, moreover 
there are Coulomb repulsions between ions); as pointed out by Kirchheim and Stolz 
long time ago \cite{kir85} an important consequence is that mobile ions behave like 
Fermions from the statistical mechanical point of view  \cite{bar99,maa99}. It is also 
clear that ionic motion takes place via transitions between different distributions of 
the mobile ions among the available sites. 

The next question is:
\begin{itemize}
\item {\it What is the nature of the individual ion sites?}
\end{itemize}
Following the commonplace assumption in most previous works we shall assume 
that the ions do not significantly perturb the network (the dynamic structure model of 
Bunde, Maass, and Ingram \cite{bun94}, of course, challenges this assumption).
Given a rigid network there are two possibilities, depending on the strength of the 
interactions between mobile ions relative to 
ion-lattice interactions. If interactions among mobile ions are relatively weak, each 
ion site has a well-defined energy $\epsilon$ which does not depend on whether or 
not neighboring sites are occupied. In this case the energy of the ionic system is 
simply the sum of all mobile ion energies. We shall initially assume that [mobile] 
ion-ion interactions are indeed relatively weak, but later on remove this limitation.

The density of ion site energies is denoted by $p(\epsilon)$. The spread of energies 
deriving from the disorder of the glassy matrix is expected to be much larger than 
$k_BT$. Consequently, to a good approximation the following picture applies: States 
up to an energy $\epsilon_F$ are filled while states above $\epsilon_F$ are empty. 
$\epsilon_F$ is the so-called Fermi energy. As is well-known from the theory of 
electronic conduction in solids \cite{kit96} there are two possibilities: 
$p(\epsilon_F)>0$ corresponds to the Fermi energy lying within an energy band 
(``metal''), while $p(\epsilon_F)=0$ corresponds to having the Fermi energy placed 
between two bands, i.e., the existence of a band gap $\Delta>0$ (``semiconductor''). 
The ``metal'' case was first treated by Kirchheim \cite{kir83} and more recently by 
Baranovskii and Cordes \cite{bar99}. We argue below that it is more realistic to 
assume the existence of a band gap. 

To be specific, consider the case of an ordinary alkali-oxide glass. The glassy 
network is created when the melt solidifies at the glass transition. The number of ions 
is equal to the number of negatively charged non-bridging oxygen (NBO) atoms. 
Because of Coulomb attraction it is favorable for each NBO atom to have at least one 
ion site associated with it. The crucial question is whether there are {\it more} low-
energy ion sites than the number of NBO atoms. Any empty low energy site is 
basically a hole in the network structure. If there were a substantial fraction of holes, 
the density of the glass would be considerably lower than the density of crystals of 
similar composition. This is never the case. We conclude that there is only one low 
energy site per mobile ion, one per NBO atom. This implies the existence of a band 
gap. 

An alternative argument for the existence of a band gap considers the annealing state 
of the glass. If the glass is well annealed, all atoms {\it including the ions} have been 
gradually and delicately brought into low-energy states defined by
surrounding atoms. It is then highly unlikely that there are more low-energy ion sites 
than the actual number of ions. Surely, the glass would have to spend energy to 
produce empty sites carefully optimized for housing an ion, energy spent
without reaping any benefits. This is like spending a lot of effort preparing for a 
guest that in the end prefers to stay elsewhere!

We have arrived at the following picture: 
\begin{itemize}
\item {\bf An ion conducting glass has ``native'' ion sites, the number of which is 
equal to the number of ions. There are also ``non-native'' ion sites in the glass, but 
these all have energies at least the band gap $\Delta$ higher than that of any
native site.}
\end{itemize}
This picture is implicit already in the 1985 paper by Kirchheim and Stolz on tracer 
diffusion and mobility of interstitials in disordered materials \cite{kir85} (cf. Fig. 
10).

Proceeding to consider the conduction mechanism, we shall refer by analogy to the 
theory of electronic conduction in semiconductors \cite{kit96}. A semiconductor has 
two sorts of charge carriers, {\it electrons} excited into the conduction band and {\it 
holes} of opposite charge (these are simply electrons missing from the valence band). 
In an intrinsic semiconductor -- the analogue of the ionic glass -- the number of 
mobile electrons is equal to the number of holes. For the ionic glass the analogue of a 
hole is a vacancy and the analogue of an excited electron is an interstitial ion, i.e., an 
ion placed in one of the high energy sites unoccupied in the ``ground state'' of the
ionic system. At any given time the number of vacancies is equal to the number of 
interstitial ions.

If $\Delta\gg k_BT$, as is assumed from here on, the number of both vacancies and 
interstitial ions is much lower than the number of mobile ions and interstitial sites. In 
this situation charge transport proceeds via motion of well-defined vacancies and 
interstitial ions. These are ``quasi-particles'' with only finite lifetime, but at the low 
quasi-particle concentrations guarantied by $k_BT\ll\Delta$ their lifetimes are long 
compared to typical jump times: Just as in semiconductors 
quasi-particles are created in pairs, move away from each other, and end their life by 
annihilating. The annihilation is a recombination where an interstitial ion jumps into 
a vacancy. In most cases the ion and vacancy annihilating are not the same as those 
originally created in a pair (note that, if they {\it are} the same, the entire process has 
not resulted in any charge transport). 

The final question is:
\begin{itemize}
\item {\it How do vacancies and interstitial ions move?}
\end{itemize}
Consider a vacancy. To move it should be filled by an ion. This ion either comes 
directly from another native site or from another native site (after one or more stops 
at interstitial sites). If $v$ is a vacancy and $i$ is an interstitial ion, the ``direct'' 
mechanism is symbolized 
\[
v\rightarrow v,
\]
while the second mechanism, because the first ion jump creates a $vi$ pair, is 
\[
v\rightarrow vvi\rightarrow ... \rightarrow vvi\rightarrow v.
\]
At low ion concentrations only the indirect mechanism is realistic.

Because of the complete symmetry between vacancies and interstitial ions we can 
immediately write up the two possible mechanisms for interstitial ion movement: 
The ``direct'' mechanism is
\[
i\rightarrow i,
\]
the ``indirect'' is 
\[
i\rightarrow iiv\rightarrow ... \rightarrow iiv\rightarrow i.
\]

So far we have assumed that interactions between mobile ions are weak, 
corresponding to low mobile ion concentration. It is likely, however, that the above 
picture applies in general: Because the glass is prepared from the liquid by gradual 
cooling, the entire ion+glass system has low energy, even for large ion 
concentrations. In contrast to the dilute case each native ion site energy now has 
substantial Coulomb contributions from neighboring mobile ions. Nevertheless, it is 
still to be expected that it takes considerable energy to move an ion out of its native 
site, simply because the entire system minimized its energy during the glass 
transition. Note the consistency of the picture: If there {\it is} a band gap, the vast 
majority of mobile ions are to be found at their native site, so the contribution to the 
native site energy from neighboring mobile ions is there basically all time.

What are the consequences of the proposed picture? Annealing a glass lowers its 
energy. One thus expects that the native ion sites lower their energy, while the energy 
of interstitial sites is expected to increase because the structure becomes more tight. 
Annealing thus increases the band gap. This implies a lowering of the conductivity, 
as always seen in experiment. Another consequence relates to our understanding of 
conductivity which is basically charge carrier density times mobility. The analogy to 
intrinsic semiconductors tells us that there are two types of charge carriers with same 
density, but not necessarily same mobility. The mobility is measured, e.g., by Hall 
effect experiments. If the vacancy mobility exceeds that of the interstitials one would 
see a sign change in the Hall effect. If vacancies and interstitials have same mobility 
there should be no Hall effect. -- Finally, we note that it is possible via correlation 
factor measurements to distinguish between vacancy and interstitial mechanism 
\cite{weg82,hei93}, in other words: determine which of the two has the largest 
mobility. For the glass of the composition ${\rm Na_2Si_2O_5}$ it is concluded that 
ion conduction proceeds via interstitials, not vacancies.

To summarize, referring to the fact that glass is produced from liquid we arrive at a 
picture of glass ion conduction as proceeding via vacancies and interstitial ions. This 
idea is not new, of course \cite{hav65,moy81,fri72}, but
has here been discussed as a direct consequence of the existence of a band gap. 
Recent computer simulations by Cormack and coworkers and by Heuer and 
coworkers \cite{cor02,heu02} are consistent with this picture. 

{\bf Conclusion:} Ionic crystals trivially have a ``native'' band gap. We suggest that 
this is also the case for ionic glasses.

\begin{acknowledgments}
This work was supported by the Danish Natural Science Research Council.
\end{acknowledgments}


\begin{thebibliography}{99}

\bibitem{ste57} J. M. Stevels, in {\it Handbuch der Physik} Vol. 20, Ed. S. 
Fl{\"u}gge (Springer, 1957), p. 350.

\bibitem{owe63} A. E. Owen, in {\it Progress in Ceramic Science}, Vol. 3 
(Macmillan, New York, 1963), p. 77.

\bibitem{hav65} Y. Haven and B. Verkerk, Phys. Chem. Glasses {\bf 6} (1965) 38.

\bibitem{isa69} J. O. Isard, J. Non-Cryst. Solids {\bf 1} (1969)  235.

\bibitem{moy81} C. T. Moynihan and A. V. Lesikar, J. Am. Ceram. Soc. {\bf 64} 
(1981) 40. 

\bibitem{ang90} C. A. Angell, Chem. Rev. {\bf 90} (1990) 523. 

\bibitem{fun90} K. Funke, Prog. Solid State Chem. {\bf 22} (1990) 111.

\bibitem{bun98} A. Bunde, K. Funke, and M. D. Ingram, Solid State Ionics {\bf 
105} (1998) 1. 

\bibitem{kir85} R. Kirchheim and U. Stolz, J. Non-Cryst. Solids {\bf 70} (1985) 
323.

\bibitem{bar99} S. D. Baranovskii and H. Cordes, J. Chem. Phys. {\bf 111} (1999) 
7546.

\bibitem{maa99} P. Maass, J. Non-Cryst. Solids {\bf 255} (1999) 35.

\bibitem{bun94} A. Bunde, M. D. Ingram, and P. Maass, J. Non-Cryst. Solids {\bf 
172-174} (1994) 1222.

\bibitem{kit96} C. Kittel, {\it Introduction to Solid State Physics}, Seventh Edition  
(Wiley, New York, 1996).
 
\bibitem{kir83} R. Kirchheim, J. Non-Cryst. Solids {\bf 55} (1983) 243.
 
\bibitem{weg82} W. Wegener and G. H. Frischat, J. Non-Cryst. Solids {\bf 50} 
(1982) 253.

\bibitem{hei93} I. Heinemann and G. H. Frischat, Phys. Chem. Glasses {\bf 34} 
(1993) 255.

\bibitem{fri72} R. J. Friauf, in {\it Physics of Electrolytes}, Vol. 1, Ed. J. Hladik 
(Academic, New York, 1972), p. 153.

\bibitem{cor02} A. N. Cormack, J. Du, and T. R. Zeitler, Phys. Chem. Chem. Phys. 
{\bf 4} (2002) 3193.

\bibitem{heu02} H. Lammert, M. Kunow, and A. Heuer, in preparation (2002). 

\end{thebibliography}
\end{document}